\documentclass[pra,twocolumn,showpacs,preprintnumbers,amsmath,amssymb]{revtex4}
\usepackage{graphicx}
\usepackage{dcolumn}
\usepackage{bm}
\usepackage{color}
\input xy
\xyoption{all}
\newtheorem{proposition}{Proposition}
\newcommand{\abs}[1]{\left| #1 \right|}
\newcommand{\ket}[1]{\left| #1 \right\rangle}
\newcommand{\acin}{Ac{\'{\i}}n }
\newcommand{\etal}{\emph{et al.}}
\newcommand{\threestd}{\chi}
\newcommand{\spacer}{\rule[0cm]{0cm}{0cm}}

\newcommand{\spc}{\hspace{5mm}}
\newcommand{\cale}{\mathcal{E}}
\newenvironment{proof}{\par\noindent\textsc{Proof.}}
                      {\nopagebreak\spacer\hfill $\square$}
\DeclareMathOperator{\hdet}{Hdet}
\DeclareMathOperator{\im}{Im}

\begin{document}

\title{Topology of the three-qubit space of entanglement types}
\author{Scott N. Walck}
  \email{walck@lvc.edu}
\author{James K. Glasbrenner}
\author{Matthew H. Lochman}
\author{Shawn A. Hilbert}
  \altaffiliation[Currently at ]{Physics Department, University of Nebraska}
\affiliation{Department of Physics, Lebanon Valley College,
             Annville, PA 17003}

\date{July 21, 2005; revised August 29, 2005}

\begin{abstract}
The three-qubit space of entanglement types is the orbit space of the
local unitary action on the space of three-qubit pure states,
and hence describes the types of entanglement that a system of three
qubits can achieve.
We show that this orbit space is homeomorphic to a certain subspace
of $\mathbb{R}^6$, which we describe completely.
We give a topologically based classification of three-qubit entanglement
types, and we argue that the nontrivial topology of the three-qubit
space of entanglement types forbids the existence of standard states
with the convenient properties of two-qubit standard states.
\end{abstract}


\maketitle

\section{Introduction}

Quantum entanglement is an important resource
in quantum computation, quantum communication, and other emerging
quantum technologies.
A fundamental problem in quantum entanglement theory is to understand
the types of entanglement that a composite quantum system can
achieve.

Linden and Popescu~\cite{linden98}
presented a program for precisely characterizing the types of entanglement
that a composite quantum system can exhibit.
In their program, the Lie group of local unitary (LU)
transformations acts on the
space of quantum states, partitioning it into orbits.
Each orbit represents a type of entanglement achievable by the
quantum system.
The collection of orbits, known as the orbit space
(see, for example, \cite{hatcher}, p. 72)
or the \emph{space of entanglement types},
then forms a mathematical object that describes all of the possible types
of entanglement for a quantum system.
In fact, Linden and Popescu regarded this orbit space as
``the main mathematical object we are investigating.''\cite{linden98}
They went on to determine the number of parameters required to
distinguish orbits in the orbit space
and to comment on the importance of
local unitary invariants in describing the orbit space.

In \cite{linden99}, Linden, Popescu, and Sudbery gave the number of
parameters needed to describe the
space of entanglement types
for mixed quantum states,
and emphasized the importance of having enough LU invariants to
separate orbits.  (A collection of invariants separates orbits
if any two states with the same values for each invariant
necessarily lie on the same orbit.  Such a collection is also
called a complete set of invariants.)

Kempe~\cite{kempe99}, Coffman \etal~\cite{coffman00}, and
Sudbery~\cite{sudbery01} identified specific
LU invariants for three-qubit pure states.

A major advance toward a complete description of the
three-qubit
space of entanglement types
came from \acin \etal~\cite{acin00,acin01}
who found a standard form for three-qubit state vectors and a convenient set
of LU invariants
that had a simple form when evaluated for standard state vectors.
The near invertibility of their expressions for invariants
allowed \acin \etal\ to argue that a set of six
LU invariants were sufficient to separate orbits, giving
unique coordinates for the orbit space.
The allowable values that the invariants could assume
remained unknown, and consequently the understanding of the
types of entanglement achievable for three qubits was incomplete.

An additional point that deserves to be emphasized
is that the space $\cale$ of entanglement types
is a topological space, and not merely
a set.
Topology provides a system for keeping
track of which types of entanglement are ``close'' to
other types.
Thus we define the \emph{space of entanglement types} for $n$
qubits, $\cale_n$, to be the orbit space
produced by the action of the local unitary group
on the $n$-qubit pure quantum state space.
In other words, $\cale_n$ is the set of LU orbits (entanglement types)
equipped with the quotient topology it inherits from the quantum state space.

The present work completes Linden and Popescu's program
for three-qubit pure states.  We precisely characterize $\cale_3$,
showing that it is homeomorphic (identical not just as a set
but as a topological space)
to a particular subspace of $\mathbb{R}^6$.

The paper is organized as follows.  In section \ref{genprog},
we give a general program for describing the
space of entanglement types for a composite
quantum system as a topological space.
In section \ref{qb2}, we apply this program
to two-qubit pure states in an effort to gain some insight in
a simple setting.  Section \ref{qb3} applies the general program
to three-qubit pure states, and constitutes the heart of the paper.
Section \ref{vissec} provides some help in visualizing the
possibilities for three-qubit entanglement types, and gives
a topologically based classification.
In section \ref{topss}, we argue that the
nontrivial topology of $\cale_3$ forbids the existence of three-qubit standard
states possessing all of the nice properties that two-qubit standard
states have.

\section{Space of Entanglement Types}
\label{genprog}
In this section, we outline a general program for describing
the space of entanglement types of a composite quantum system
as a topological imbedding
(see, for example, \cite{munkres}, p. 105) in $\mathbb{R}^m$ for some $m$.

Let $H = \mathbb{C}^d$ denote the Hilbert space of a quantum system.
Quantum states correspond to rays in the Hilbert space, so
the space of quantum states is
$\mathbb{P}(\mathbb{C}^d) = \mathbb{CP}^{d-1}$,
the $(d-1)$-dimensional complex projective space.
The $(2d-1)$-sphere, $S^{2d-1}$, is the collection of normalized
state vectors in $\mathbb{R}^{2d} \approx \mathbb{C}^d$.
Let $T$ be the space on which the group of LU transformations acts.
If the group $G$ of LU transformations
contains operations that change the phase of state vectors
(as the groups $U(2) \times U(2) \times U(2)$ and
$SU(2) \times SU(2) \times SU(2) \times U(1)$ do for three qubits),
then it does not matter whether the orbit space is formed from the space
of quantum states or from the space of normalized quantum state vectors.
In the notation of \cite{hatcher}, the definition of the space of
entanglement types is
\[
\cale \equiv T/G = \mathbb{CP}^{d-1}/G = S^{2d-1}/G .
\]
In the sequel, we require the space $T$ to be topologically compact.
Both $\mathbb{CP}^{d-1}$ and $S^{2d-1}$ are compact, and both lead
to the correct orbit space, so either may be chosen for the space $T$.
For mixed states, one could take $T$ to be the space of density
matrices.

\begin{proposition}
\label{genprogprop}
Let $(I_1,\dots,I_m)$
be a complete set of continuous, real-valued invariants.
Applying the collection of invariants to
a quantum state provides a continuous map $I: T \to \mathbb{R}^m$.
If $X \subset \mathbb{R}^m$ is the image of this map,
then the space $\cale$ of entanglement types is homeomorphic to $X$.
\end{proposition}
\begin{proof}
Since the map $I$ is constructed from LU invariants, it must factor
through the space $\cale$ of entanglement types:  $I = f \circ \pi$,
where $\pi$ is the projection map that associates a quantum state
with its entanglement type in $\cale$.
(The map $f$ is well defined because $I$ is constructed from LU invariants.)
\[
\xymatrix{
T \ar[d]_\pi \ar[dr]^I & \\
\cale \ar[r]_f & X
}
\]
Since there are enough LU invariants to separate orbits, the map
$f$ is one-to-one, and since $X$ is the image of the map $I$,
the map $f$ is onto.
The map $\pi$ is continuous since it defines the quotient topology
on $\cale$, and $I$ is continuous because its components are continuous,
so $f$ is continuous by a standard result about quotient spaces
(see \cite{munkres}, Theorem 22.2).
The map $f: \cale \to X$ is then a continuous bijection.
Since $\cale$ is compact (it is the image of the compact space $T$
under the continuous map $\pi$) and $X$ is Hausdorff (a subspace of
a Hausdorff space is Hausdorff),
we have that $f$ is a homeomorphism.
(See, for example, \cite{munkres}, Theorem 26.6.)
\end{proof}

\section{Description of $\cale_2$}
\label{qb2}
In this section, we apply the procedure of the previous section to
pure states of two qubits to see how it works in a familiar
setting.  For pure states of two qubits, the Hilbert space is
$\mathbb{C}^2 \otimes \mathbb{C}^2 \approx \mathbb{C}^4$,
the quantum state space is
$\mathbb{P}(\mathbb{C}^2 \otimes \mathbb{C}^2) \approx \mathbb{CP}^3$,
and the space
of normalized quantum state vectors is $S^7$.  The group of local unitary
transformations is $U(2) \times U(2)$.
The space of entanglement types is
\[
\cale_2 = \mathbb{P}(\mathbb{C}^2 \otimes \mathbb{C}^2)/U(2) \times U(2)
  = S^7/U(2) \times U(2) .
\]

A single LU invariant is all that is required to separate orbits
for pure states of two qubits.
For a normalized two-qubit pure state
\[
\ket{\psi} = \sum_{i=0}^1 \sum_{j=0}^1 \psi_{ij} \ket{ij} ,
\]
the concurrence introduced by
Hill and Wootters~\cite{hill97},
\[
C(\psi) = 2 \abs{\psi_{00} \psi_{11} - \psi_{01} \psi_{10}} ,
\]
is a continuous, real-valued invariant.
The Schmidt decomposition theorem~\cite{ekert95,aravind96}
tells us that this single
invariant is sufficient to separate orbits, so it suffices to use
$m = 1$ invariant for two qubits.
It is easily shown
that the image of the concurrence map is
$[0,1] \subset \mathbb{R}^1$, the closed interval
from zero to one.
We conclude that $\cale_2$ is homeomorphic to $[0,1]$.
Two-qubit entanglement types are in one-to-one correspondence
with the closed unit interval, with $0$ representing unentangled
states, $1$ representing fully entangled states, and numbers in
the open unit interval representing varying degrees of partially
entangled states.

Table~\ref{tabtypes2qb} gives a topologically based classification
of two-qubit entanglement types,
provided mainly for later comparison with Table~\ref{tabtypes}
which classifies three-qubit entanglement types.
In this table, we view $\cale_2$ as a cell complex (or CW complex)
\cite{hatcher} composed of two 0-cells (points) and one 1-cell
(open unit interval).
One 0-cell, $e^0_{SEP}$, represents the separable (unentangled) states.
The other 0-cell, $e^0_{EPR}$, represents fully entangled states,
such as the EPR pair, $\ket{EPR} = 1/\sqrt{2}(\ket{01} - \ket{10})$.
The 1-cell, $e^1$, representing all of the partial entanglement types
is attached to $e^0_{SEP}$ at one end and to $e^0_{EPR}$ at the other end.

Entanglement types considered in the present work are more precisely
called LU entanglement types, since each type is associated with
an equivalence class of quantum states under local unitary transformation.
An alternative equivalence relation on the space of quantum states is
is equivalence under stochastic local operations
and classical communication (SLOCC)~\cite{dur00}.
Each SLOCC equivalence class of quantum states can be regarded as
an SLOCC entanglement type in the same way that
each LU equivalence class of quantum states is regarded as
an LU entanglement type.
For two-qubit pure states, there are only two SLOCC classes:
unentangled and entangled.
LU operations are a subset of SLOCC operations,
so each LU entanglement type that we have identified is associated
with exactly one SLOCC entanglement type.
In Table~\ref{tabtypes2qb}, we list the SLOCC class
associated with each collection of LU entanglement types.

Each LU entanglement type corresponds to an LU orbit
(an equivalence class) of quantum states.
This LU orbit is a differentiable manifold of quantum states.
The fourth column of Table~\ref{tabtypes2qb} lists the dimension
of the LU orbit of quantum states for each entanglement
type~\cite{kus01,walckICQI}.

\begin{table}
\caption{\label{tabtypes2qb}
Two-qubit space of entanglement types viewed as a cell complex.
The superscript in the cell name gives the dimension of the cell.
}
\begin{ruledtabular}
\begin{tabular}{cccc}
Cell        &                   & SLOCC       & Orbit     \\
name        & Concurrence       & Class       & Dimension \\ \hline
$e^0_{SEP}$ & $C(\psi) = 0$     & unentangled & 4         \\
$e^1$       & $0 < C(\psi) < 1$ &   entangled & 5         \\
$e^0_{EPR}$ & $C(\psi) = 1$     &   entangled & 3
\end{tabular}
\end{ruledtabular}
\end{table}

\section{Description of $\cale_3$}
\label{qb3}
In this section, we apply the procedure of section \ref{genprog} to
pure states of three qubits.
For pure states of three qubits, the Hilbert space is
$\mathbb{C}^2 \otimes \mathbb{C}^2 \otimes \mathbb{C}^2 \approx \mathbb{C}^8$,
the quantum state space is
$\mathbb{P}(\mathbb{C}^2 \otimes \mathbb{C}^2 \otimes \mathbb{C}^2)
  \approx \mathbb{CP}^7$, and the space
of normalized quantum state vectors is $S^{15}$.  The group of local unitary
transformations is $U(2) \times U(2) \times U(2)$.
The space of entanglement types is
\begin{align*}
\cale_3 &= \mathbb{P}(\mathbb{C}^2 \otimes \mathbb{C}^2 \otimes \mathbb{C}^2)
  / U(2) \times U(2) \times U(2) \\
        &= S^{15} / U(2) \times U(2) \times U(2) .
\end{align*}
Let
\[
\ket{\psi} = \sum_{i=0}^1 \sum_{j=0}^1 \sum_{k=0}^1 \psi_{ijk} \ket{ijk}
\]
denote a normalized three-qubit state vector.
We will use the symbol $\psi$ without ket notation to denote
the column vector of coefficients $\psi_{ijk}$ in the standard basis.

In order to apply the program of section \ref{genprog}, our first
step is to choose some LU invariants.
Let $n$ be a positive integer, and let $\sigma$ and $\tau$ be permutations
on $n$ elements.
For each $n$, $\sigma$, and $\tau$,
define a function
$P_{\sigma,\tau}^n: \mathbb{C}^8 \to \mathbb{C}$ by
\begin{widetext}
\begin{equation}
\label{pform}
P_{\sigma,\tau}^n(\psi)
 = \sum_{i_1=0}^1 \sum_{j_1=0}^1 \sum_{k_1=0}^1
             \cdots \sum_{i_n=0}^1 \sum_{j_n=0}^1 \sum_{k_n=0}^1
  \psi_{i_1 j_1 k_1} \cdots \psi_{i_n j_n k_n}
  \psi_{i_1 j_{\sigma(1)} k_{\tau(1)}}^* \cdots
  \psi_{i_n j_{\sigma(n)} k_{\tau(n)}}^*
\end{equation}
\end{widetext}
Every function of this form is an LU invariant
\cite{sudbery01,weyl,gingrich02}.
Define
\begin{align*}
I_1(\psi) &= P_{e,(12)}^2(\psi) \\
I_2(\psi) &= P_{(12),e}^2(\psi) \\
I_3(\psi) &= P_{(12),(12)}^2(\psi) \\
I_4(\psi) &= P_{(123),(132)}^3(\psi) \\
I_5(\psi) &= \abs{\hdet(\psi)}^2 \\
I_6(\psi) &= \im[P_{(34)(56),(13524)}^6(\psi)]
\end{align*}
Invariants $I_1$ through $I_4$ match those given in~\cite{gingrich02}.
It is straightforward to show that they are real-valued invariants.
$\hdet$ is Cayley's hyperdeterminant used in~\cite{acin00}
and corresponding to the three-tangle of
\cite{coffman00,sudbery01}.

The authors of~\cite{acin00,acin01,gingrich02} chose a discrete invariant
as a sixth invariant in an effort to avoid redundancy.
Because we are interested in topology
and we want to preserve the information about which orbits are close
to other orbits, we choose $I_6$
to be a continuous invariant rather than a discrete one.

\acin and coworkers~\cite{acin00,acin01}
introduced a new set of invariants that made it
possible to argue that the invariants separate orbits.
\begin{align}
\label{j1def}
J_1(\psi) &= \frac{1}{4} \left[ 1 - I_1(\psi) - I_2(\psi) + I_3(\psi)
               - 2 \sqrt{I_5(\psi)} \right] \\
\label{j2def}
J_2(\psi) &= \frac{1}{4} \left[ 1 - I_1(\psi) + I_2(\psi) - I_3(\psi)
               - 2 \sqrt{I_5(\psi)} \right] \\
\label{j3def}
J_3(\psi) &= \frac{1}{4} \left[ 1 + I_1(\psi) - I_2(\psi) - I_3(\psi)
               - 2 \sqrt{I_5(\psi)} \right] \\
\label{j4def}
J_4(\psi) &= \sqrt{I_5(\psi)} \\
J_5(\psi) &= \frac{5}{12}
   - \frac{1}{4} I_1(\psi) - \frac{1}{4} I_2(\psi) - \frac{1}{4} I_3(\psi)
     \notag \\
\label{j5def}
   & \spc + \frac{1}{3} I_4(\psi) - \frac{1}{2} \sqrt{I_5(\psi)} \\
\label{j6def}
J_6(\psi) &= I_6(\psi)
\end{align}
Invariants $J_1$ through $J_5$ match those given in both
\cite{acin00,acin01} and~\cite{gingrich02}.

The invariants $J_i$ have nice properties under qubit permutation.
Invariant $J_1$, which reports on the two-qubit entanglement
between qubits 2 and 3, is invariant under interchange of qubits 2 and 3.
Similarly, $J_2$ ($J_3$) is invariant under interchange of qubits
1 and 3 (1 and 2).
LU invariants $J_4$, $J_5$, and $J_6$
encode the essential three-qubit entanglement
of the state, and
remain unchanged under any permutation of qubits.

\acin \etal~\cite{acin00,acin01}
generalized the Schmidt decomposition to three qubits,
showing that every normalized three-qubit state vector
is LU-equivalent to a standard state vector of the form
\begin{equation}
\label{stdform}
\ket{\threestd} = \lambda_0 \ket{000}
 + \lambda_1 e^{i \phi} \ket{100}
 + \lambda_2 \ket{101}
 + \lambda_3 \ket{110}
 + \lambda_4 \ket{111} ,
\end{equation}
where the $\lambda_i$ are real and nonnegative, $\phi$ is real, and
$\sum_{i=0}^4 \lambda_i^2 = 1$.
Evaluating invariants (\ref{j1def})--(\ref{j6def}) for the standard
state vector $\threestd$, we have~\cite{acin00,acin01,gingrich02}
\begin{align}
\label{j1}
J_1(\threestd) &=
  \abs{\lambda_1 \lambda_4 e^{i \phi} - \lambda_2 \lambda_3}^2 \\
\label{j2}
J_2(\threestd) &= \lambda_0^2 \lambda_2^2 \\
\label{j3}
J_3(\threestd) &= \lambda_0^2 \lambda_3^2 \\
\label{j4}
J_4(\threestd) &= \lambda_0^2 \lambda_4^2 \\
\label{j5}
J_5(\threestd) &= 2 \lambda_0^2 \lambda_2^2 \lambda_3^2
         - 2 \lambda_0^2 \lambda_1 \lambda_2 \lambda_3 \lambda_4 \cos \phi \\
J_6(\threestd) &= \lambda_0^4 \lambda_1 \lambda_2 \lambda_3 \lambda_4
  \sin \phi \notag \\
\label{j6}
  & \spc \times (2 \lambda_0^2 \lambda_4^2 + 2 \lambda_1^2 \lambda_4^2
       - \lambda_4^2 - 2 \lambda_1 \lambda_2 \lambda_3 \lambda_4 \cos \phi) .
\end{align}

Invariants $J_1$ through $J_6$ are a complete set of continuous,
real-valued invariants.
If we can find the image in $\mathbb{R}^6$ that these invariants
make, then by Proposition \ref{genprogprop}, we can completely
characterize the space of entanglement types for three qubits.
The next proposition does this.

\begin{proposition}
\label{mainprop}
The three-qubit space of entanglement types, $\cale_3$, is
homeomorphic to the subspace of $\mathbb{R}^6$ consisting
of points
$(\beta_1,\beta_2,\beta_3,\beta_4,\beta_5,\beta_6) \in \mathbb{R}^6$
that satisfy
\begin{align}
\label{beta1range}
0 & \leq \beta_1 \leq \frac{1}{4}, \\
\label{beta2range}
0 & \leq \beta_2 \leq \frac{1}{4}, \\
\label{beta3range}
0 & \leq \beta_3 \leq \frac{1}{4}, \\
\label{beta4range}
0 & \leq \beta_4 \leq \frac{1}{4},
\end{align}
\begin{equation}
\label{lastfound}
\beta_1 \beta_2 + \beta_1 \beta_3 + \beta_2 \beta_3 +
(\beta_1 + \beta_2 + \beta_3) \beta_4 +
\beta_4^2 \leq
\frac{1}{4} \beta_4 + \frac{1}{2} \beta_5 ,
\end{equation}
and
\begin{multline}
\label{beta6}
\left[ (\beta_5 + \beta_4)^2
    - 4 (\beta_1 + \beta_4)(\beta_2 + \beta_4)(\beta_3 + \beta_4) \right] \\
\times
\left[ \beta_5^2 - 4 \beta_1 \beta_2 \beta_3 \right]
 + 4 \beta_6^2 = 0 .
\end{multline}
\end{proposition}
\begin{proof}
Define a map $J: S^{15} \to \mathbb{R}^6$ by
\[
J(\psi) = \left( J_1(\psi),J_2(\psi),J_3(\psi),J_4(\psi),J_5(\psi),J_6(\psi)
          \right) .
\]
Let $X$ be the subspace of $\mathbb{R}^6$ satisfying conditions
(\ref{beta1range})--(\ref{beta6}) above.

We claim that $X$ is the image of the map $J$.
From this and Proposition \ref{genprogprop} we may conclude that
$X$ is homeomorphic to $\cale_3$.
We will show first
that $J$ maps into $X$ (so that $X$ contains the image of $J$),
and then that $J$ maps onto $X$ (so that $X$ is contained in the image of $J$).

Before we proceed with main part of the proof,
we wish to note some conditions
that are implied by conditions (\ref{beta1range})--(\ref{beta6}) above.
\begin{equation}
\label{bffgez}
\beta_4 + \beta_5 \geq 0
\end{equation}
\begin{equation}
\label{bottomsurface}
\beta_5^2 - 4 \beta_1 \beta_2 \beta_3 \leq 0
\end{equation}
\begin{equation}
\label{deltabetacond}
\Delta_\beta \equiv (\beta_5 + \beta_4)^2
    - 4 (\beta_1 + \beta_4)(\beta_2 + \beta_4)(\beta_3 + \beta_4) \geq 0
\end{equation}
Condition (\ref{bffgez}) follows from (\ref{lastfound}) and the fact
that $\beta_1$, $\beta_2$, $\beta_3$, and $\beta_4$ are nonnegative.
To establish (\ref{bottomsurface}), assume the contrary, that
$\beta_5^2 - 4 \beta_1 \beta_2 \beta_3 > 0$.
Equation (\ref{beta6}) then requires that
$4 (\beta_1 + \beta_4)(\beta_2 + \beta_4)(\beta_3 + \beta_4)
   - (\beta_5 + \beta_4)^2 \geq 0$,
and adding these two inequalities violates (\ref{lastfound}).
Hence (\ref{bottomsurface}) is established.
To establish (\ref{deltabetacond}), assume the contrary, that
$4 (\beta_1 + \beta_4)(\beta_2 + \beta_4)(\beta_3 + \beta_4)
   - (\beta_5 + \beta_4)^2 > 0$.
Equation (\ref{beta6}) then requires that
$\beta_5^2 - 4 \beta_1 \beta_2 \beta_3 \geq 0$,
and adding these two inequalities violates (\ref{lastfound}).
Hence (\ref{deltabetacond}) is established.

To show that $J$ maps into $X$, we must show that
$J(\psi) \in X$ for all normalized three-qubit state vectors $\psi$.
In other words, if
$\beta = J(\psi)$ for a normalized three-qubit state vector
$\psi$, then
$\beta = (\beta_1,\beta_2,\beta_3,\beta_4,\beta_5,\beta_6)$
satisfies conditions (\ref{beta1range})--(\ref{beta6}) above.

In~\cite{acin01}, \acin \etal\ record conditions
(\ref{beta1range})--(\ref{beta4range})
as well as (\ref{deltabetacond}),
so we will assume these and move on to prove
(\ref{lastfound}) and (\ref{beta6}).
Let $\beta = J(\psi)$ for a normalized three-qubit state vector
$\psi$.  Let $\threestd$ be a three-qubit state vector in standard form
(\ref{stdform}) that is LU-equivalent to $\psi$.  Then
$\beta = J(\threestd)$.

Let us prove (\ref{lastfound}) for $\beta = J(\psi)$.
Our approach is to prove (\ref{bottomsurface}), and then to use
(\ref{bottomsurface}) and (\ref{deltabetacond}) to prove (\ref{lastfound}).
For condition (\ref{bottomsurface}), equations
(\ref{j1}), (\ref{j2}), (\ref{j3}), and (\ref{j5}) give
\begin{equation}
\label{bottomstd}
\beta_1 \beta_2 \beta_3 - \frac{\beta_5^2}{4}
 = \lambda_0^4 \lambda_1^2 \lambda_2^2 \lambda_3^2 \lambda_4^2 \sin^2 \phi
 \geq 0 .
\end{equation}
If $\beta_4 > 0$, then (\ref{lastfound}) is obtained by subtracting
(\ref{bottomsurface}) from (\ref{deltabetacond}), and dividing by
$\beta_4$.
If $\beta_4 = 0$, we must show that
\begin{equation}
\label{lastsimple}
\beta_1 \beta_2 + \beta_1 \beta_3 + \beta_2 \beta_3 \leq
\frac{1}{2} \beta_5 .
\end{equation}
By equation (\ref{j4}), either $\lambda_0 = 0$ or $\lambda_4 = 0$.
If $\lambda_0 = 0$, then
equations (\ref{j2}), (\ref{j3}), and (\ref{j5}) show that
$\beta_2 = \beta_3 = \beta_5 = 0$, and
(\ref{lastsimple}) is satisfied.
Alternatively, if $\lambda_4 = 0$, then equations
(\ref{j1}), (\ref{j2}), (\ref{j3}), and (\ref{j5}) give
\[
\beta_1 \beta_2 + \beta_1 \beta_3 + \beta_2 \beta_3
 = \lambda_0^2 \lambda_2^2 \lambda_3^2
   (\lambda_0^2 + \lambda_2^2 + \lambda_3^2) \leq \frac{1}{2} \beta_5 .
\]
This completes the proof of condition (\ref{lastfound}) for
$\beta = J(\psi)$.

For condition (\ref{beta6}),
note from equations (\ref{j1})--(\ref{j5})
that
\begin{multline}
(\beta_4 + \beta_5)^2
    - 4 (\beta_1 + \beta_4)(\beta_2 + \beta_4)(\beta_3 + \beta_4) \\
  = \lambda_0^4
  (2 \lambda_0^2 \lambda_4^2 + 2 \lambda_1^2 \lambda_4^2
       - \lambda_4^2 - 2 \lambda_1 \lambda_2 \lambda_3 \lambda_4 \cos \phi)^2 .
\end{multline}
This, along with equations (\ref{bottomstd}) and (\ref{j6}) gives
condition (\ref{beta6}).
This completes the proof for the claim that $J$ maps into $X$.

Next we address the claim that $J$ maps onto $X$.
Here we must show for any $\beta \in X$ that there exists a
normalized three-qubit state vector $\psi$ such that $J(\psi) = \beta$.
In fact we will produce a normalized three-qubit state vector
$\threestd$ in standard form (\ref{stdform}) such that
$J(\threestd) = \beta$.

\begin{trivlist}
\item \textbf{Case 1:}  $\beta_1 + \beta_4 = 0$.
In this case, conditions (\ref{beta1range}) and (\ref{beta4range})
imply that $\beta_1 = 0$ and $\beta_4 = 0$.
Condition (\ref{bottomsurface}) then implies that $\beta_5 = 0$,
condition (\ref{beta6}) implies that $\beta_6 = 0$,
and conditions (\ref{beta2range}), (\ref{beta3range}), and
(\ref{lastfound}) imply that $\beta_2 \beta_3 = 0$.
\item \textbf{Case 1a:}  $\beta_2 = 0$.
Let
\begin{equation}
\ket{\threestd} = \sqrt{\frac{1}{2} + \sqrt{\frac{1}{4} - \beta_3}} \ket{000}
                + \sqrt{\frac{1}{2} - \sqrt{\frac{1}{4} - \beta_3}} \ket{110} .
\end{equation}
Notice that condition (\ref{beta3range}) guarantees real nonnegative entries
for $\threestd$.
\item \textbf{Case 1b:}  $\beta_3 = 0$.
Let
\begin{equation}
\ket{\threestd} = \sqrt{\frac{1}{2} + \sqrt{\frac{1}{4} - \beta_2}} \ket{000}
                + \sqrt{\frac{1}{2} - \sqrt{\frac{1}{4} - \beta_2}} \ket{101} .
\end{equation}
Notice that condition (\ref{beta2range}) guarantees real nonnegative entries
for $\threestd$.

\item \textbf{Case 2:}  $\beta_1 + \beta_4 > 0$.
Let
\begin{equation}
\label{lam0def}
\lambda_0^2 = \frac{\beta_4 + \beta_5 + \sqrt{\Delta_\beta}}
                   {2 (\beta_1 + \beta_4)} .
\end{equation}
Conditions (\ref{bffgez}) and (\ref{deltabetacond}) imply that
$\lambda_0^2$ is nonnegative.
Choose $\lambda_0$ to be nonnegative.

\item \textbf{Case 2a:}  $\lambda_0 = 0$.
Condition (\ref{bffgez}) implies that $\beta_4 + \beta_5 = 0$.
Condition (\ref{deltabetacond}), along with conditions
(\ref{beta1range}), (\ref{beta2range}), (\ref{beta3range}),
and (\ref{beta4range}),
then implies that $\beta_4 = 0$ (and hence that $\beta_5 = 0$),
and also that $\beta_1 \beta_2 \beta_3 = 0$.
Condition (\ref{beta6}) implies that $\beta_6 = 0$.
We must have $\beta_1 > 0$, and then condition (\ref{lastfound})
implies $\beta_2 = \beta_3 = 0$.
Let
\begin{equation}
\ket{\threestd} = \sqrt{\frac{1}{2} + \sqrt{\frac{1}{4} - \beta_1}} \ket{100}
                + \sqrt{\frac{1}{2} - \sqrt{\frac{1}{4} - \beta_1}} \ket{111} .
\end{equation}
Notice that condition (\ref{beta1range}) guarantees real nonnegative entries
for $\threestd$.

\item \textbf{Case 2b:}  $\lambda_0 > 0$.
Let
\[
\lambda_2 = \frac{\sqrt{\beta_2}}{\lambda_0} , \spc
\lambda_3 = \frac{\sqrt{\beta_3}}{\lambda_0} , \spc
\lambda_4 = \frac{\sqrt{\beta_4}}{\lambda_0} .
\]

\item \textbf{Case 2b1:}  $\beta_4 = 0$.
Since Case 2 requires $\beta_1 + \beta_4 > 0$, we have
$\beta_1 > 0$.
Conditions (\ref{bottomsurface}) and (\ref{deltabetacond})
imply that $\Delta_\beta = 0$, and hence that
\[
\lambda_0^2 = \frac{\beta_5}{2 \beta_1} .
\]
Since Case 2b requires $\lambda_0 > 0$, we must have $\beta_5 > 0$.
Let
\begin{multline}
\ket{\threestd} = \sqrt{\frac{\beta_5}{2 \beta_1}} \ket{000}
 + \sqrt{1-\frac{\beta_5}{2 \beta_1}
                 -\frac{2 \beta_1 \beta_2}{\beta_5}
                 -\frac{2 \beta_1 \beta_3}{\beta_5}} \ket{100} \\
 + \sqrt{\frac{2 \beta_1 \beta_2}{\beta_5}} \ket{101}
 + \sqrt{\frac{2 \beta_1 \beta_3}{\beta_5}} \ket{110} .
\end{multline}
Also note that when $\beta_4 = 0$,
conditions (\ref{bottomsurface}) and (\ref{deltabetacond})
imply that
\begin{equation}
\label{b5123eq}
\beta_5 = 2 \sqrt{\beta_1 \beta_2 \beta_3} .
\end{equation}
Condition (\ref{beta6}) then implies that $\beta_6 = 0$.
When $\beta_4 = 0$,
condition (\ref{lastfound}) becomes (\ref{lastsimple}).
Multiplying the latter by $2/\beta_5$
and using (\ref{b5123eq}), we get
\[
\frac{2 \beta_1 \beta_2}{\beta_5} + \frac{2 \beta_1 \beta_3}{\beta_5}
 + \frac{\beta_5}{2 \beta_1} \leq 1 ,
\]
which shows that all of the entries in $\threestd$ above are
real and nonnegative.

\item \textbf{Case 2b2:}  $\beta_4 > 0$, $\beta_2 = 0$.
Condition (\ref{bottomsurface}) implies that $\beta_5 = 0$.
Condition (\ref{beta6}) implies that $\beta_6 = 0$.
Let
\[
\ket{\threestd} = \lambda_0 \ket{000}
  + \lambda_0 \sqrt{\frac{\beta_1}{\beta_4}} \ket{100}
  + \frac{\sqrt{\beta_3}}{\lambda_0} \ket{110}
  + \frac{\sqrt{\beta_4}}{\lambda_0} \ket{111} ,
\]
where $\lambda_0$ is given by (\ref{lam0def}).
The entries in $\threestd$ are clearly nonnegative.

\item \textbf{Case 2b3:}  $\beta_4 > 0$, $\beta_3 = 0$.
Condition (\ref{bottomsurface}) implies that $\beta_5 = 0$.
Condition (\ref{beta6}) implies that $\beta_6 = 0$.
Let
\[
\ket{\threestd} = \lambda_0 \ket{000}
 + \lambda_0 \sqrt{\frac{\beta_1}{\beta_4}} \ket{100}
 + \frac{\sqrt{\beta_2}}{\lambda_0} \ket{101}
 + \frac{\sqrt{\beta_4}}{\lambda_0} \ket{111} ,
\]
where $\lambda_0$ is given by (\ref{lam0def}).
The entries in $\threestd$ are clearly nonnegative.

\item \textbf{Case 2b4:}
$\beta_2 \beta_3 \beta_4 > 0$,
$\Delta_\beta = 0$.
Condition (\ref{beta6}) implies that $\beta_6 = 0$.
Let
\begin{multline}
\ket{\threestd} = \lambda_0 \ket{000} \\
 + \left( \frac{2 \beta_2 \beta_3 - \lambda_0^2 \beta_5}
               {2 \lambda_0 \sqrt{\beta_2 \beta_3 \beta_4}}
          + i \lambda_0 \frac{\sqrt{\beta_1 \beta_2 \beta_3 - \beta_5^2/4}}
               {\sqrt{\beta_2 \beta_3 \beta_4}} \right) \ket{100} \\
 + \frac{\sqrt{\beta_2}}{\lambda_0} \ket{101}
 + \frac{\sqrt{\beta_3}}{\lambda_0} \ket{110}
 + \frac{\sqrt{\beta_4}}{\lambda_0} \ket{111} ,
\end{multline}
where $\lambda_0$ is given by (\ref{lam0def}).
The entries are all well defined since $\lambda_0 > 0$ in Case 2b.
Note also that $\lambda_0$, $\lambda_2$, $\lambda_3$, and $\lambda_4$
are nonnegative.

\item \textbf{Case 2b5:}  $\beta_2 \beta_3 \beta_4 > 0$, $\Delta_\beta > 0$.
Let
\begin{multline}
\ket{\threestd} = \lambda_0 \ket{000} \\
 + \left( \frac{2 \beta_2 \beta_3 - \lambda_0^2 \beta_5}
               {2 \lambda_0 \sqrt{\beta_2 \beta_3 \beta_4}}
       + i \lambda_0 \frac{\beta_6}
           {\sqrt{\beta_2 \beta_3 \beta_4 \Delta_\beta}} \right) \ket{100} \\
 + \frac{\sqrt{\beta_2}}{\lambda_0} \ket{101}
 + \frac{\sqrt{\beta_3}}{\lambda_0} \ket{110}
 + \frac{\sqrt{\beta_4}}{\lambda_0} \ket{111} ,
\end{multline}
where $\lambda_0$ is given by (\ref{lam0def}).
The entries are all well defined since $\lambda_0 > 0$ in Case 2b.
Note also that $\lambda_0$, $\lambda_2$, $\lambda_3$, and $\lambda_4$
are nonnegative.
\end{trivlist}
In each case,
a bit of algebra shows that $\threestd$ is normalized.
Using (\ref{j1})--(\ref{j6}), one can show that
$J(\threestd) = \beta$.
\end{proof}

\section{Visualization of $\cale_3$}
\label{vissec}

The description of $\cale_3$ given by Proposition \ref{mainprop}
is precise and complete; nevertheless we want a way to visualize
the possibilities for three-qubit entanglement.
One way to proceed is to create a hierarchy of the six invariants
we used in describing orbits.
At the top of our hierarchy we choose $\beta_4$.
Our $\beta_4$ is related to the
3-tangle of~\cite{coffman00}, and controls the amount of ``GHZ entanglement''
possessed by a state.  If any qubit is unentangled from the other two,
then $\beta_4 = 0$, and if $\beta_4 = 1/4$, its maximum value, the
three qubits have the type of entanglement possessed by the GHZ state,
$\ket{GHZ} = 1/\sqrt{2} (\ket{000} + \ket{111})$. 

\begin{figure}
\scalebox{0.5}{\includegraphics{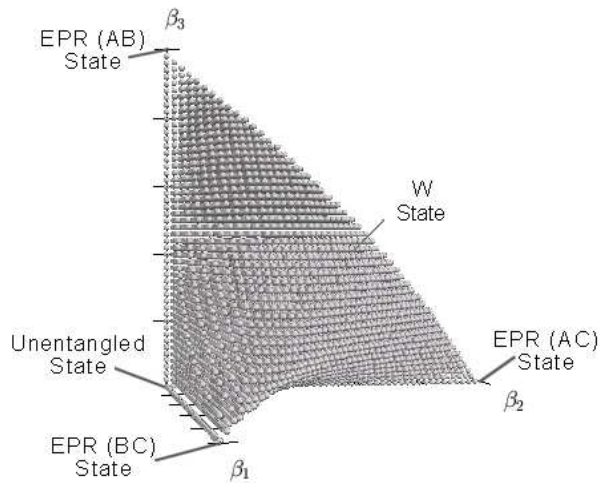}}
\caption{\label{b4zfig}Entanglement types for $\beta_4 = 0$.}
\end{figure}

We first consider the entanglement possibilities for $\beta_4 = 0$.
When $\beta_4 = 0$, conditions (\ref{bffgez}), (\ref{bottomsurface}),
and (\ref{deltabetacond}) imply that
$\beta_5 = 2 \sqrt{\beta_1 \beta_2 \beta_3}$ and
condition (\ref{beta6}) then implies that $\beta_6 = 0$.
Condition (\ref{lastfound}) becomes
\begin{equation}
\label{b123b4z}
\beta_1 \beta_2 + \beta_1 \beta_3 + \beta_2 \beta_3 \leq
  \sqrt{\beta_1 \beta_2 \beta_3} .
\end{equation}
The allowable values for $\beta_1$, $\beta_2$, and $\beta_3$ are
given by (\ref{beta1range}), (\ref{beta2range}), (\ref{beta3range}), and
(\ref{b123b4z}).
The entanglement types when $\beta_4 = 0$ are shown in
Figure~\ref{b4zfig}.  They fill a region in
$\beta_1$,$\beta_2$,$\beta_3$ space bounded by three closed
line segments $[0,1/4]$ and the bubble-like surface in which
(\ref{b123b4z}) attains equality.  Each point in Figure
\ref{b4zfig} represents a type of entanglement for three qubits
in which $\beta_4 = 0$ (no GHZ-like entanglement).
The closed line segments $[0,1/4]$ along each axis are copies
of $\cale_2$, the space of entanglement types for two qubits.
Points on these axes represent types of entanglement
in which one qubit is unentangled from the other two.
Completely unentangled states are represented by the point at the origin
in the figure.
The point in the center of the bubble-like surface represents the
type of entanglement of the W state,
$\ket{W} = 1/\sqrt{3}(\ket{100}+\ket{010}+\ket{001})$.
Since $\beta_5$ and $\beta_6$ are determined by
$\beta_1$, $\beta_2$, and $\beta_3$, Figure~\ref{b4zfig}
is a complete picture of the types of three-qubit entanglement
when $\beta_4 = 0$.

\begin{figure}
\scalebox{0.5}{\includegraphics{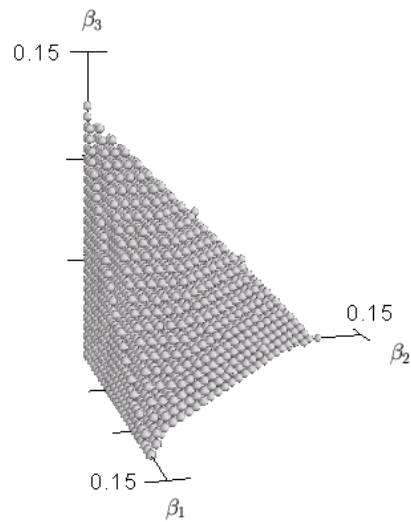}}
\caption{\label{b4midfig}Entanglement possibilities for $\beta_4 = 1/8$.
Each point in the interior of the deformed tetrahedron
has a circle of entanglement types in the $\beta_5$,$\beta_6$ plane,
as shown in Figure~\ref{b56fig}.
Each point on the boundary of the tetrahedron represents a single
entanglement type.}
\end{figure}

\begin{figure}
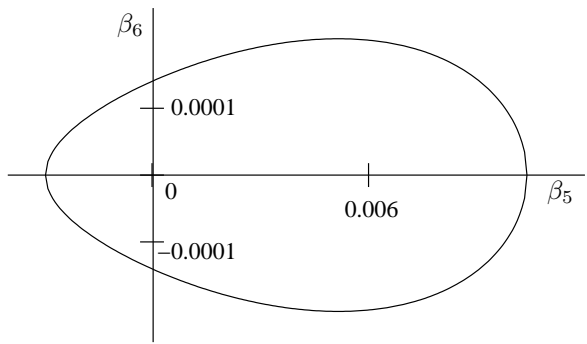
\caption{\label{b56fig}
Each point on this deformed circle is an entanglement type with
$\beta_4 = 1/8$, $\beta_1 = \beta_2 = \beta_3 = 0.03$.
Every point
in the interior of the deformed tetrahedron of Figure~\ref{b4midfig}
has a circle of entanglement types like this.
Points on the boundary of Figure~\ref{b4midfig} have a single entanglement
type.}
\end{figure}

Next we consider the entanglement possibilities for
a fixed value of $\beta_4$ in the range $0 < \beta_4 < 1/4$.
In this case, the conditions of Proposition \ref{mainprop}
produce the following allowable values of $\beta_1$, $\beta_2$, $\beta_3$.
\begin{equation}
\beta_1 \geq 0 , \spc
\beta_2 \geq 0 , \spc
\beta_3 \geq 0
\end{equation}
\begin{multline}
\label{fdef}
F \equiv \beta_4 \left( \frac{1}{4} - \beta_4 \right) 
  + \sqrt{\beta_1 \beta_2 \beta_3}
  - \beta_1 \beta_2 - \beta_1 \beta_3 - \beta_2 \beta_3 \\
  - (\beta_1 + \beta_2 + \beta_3) \beta_4 \geq 0
\end{multline}
This region in $\beta_1$,$\beta_2$,$\beta_3$ space is a deformed
tetrahedron bounded by four surfaces, as shown in Figure~\ref{b4midfig}.
For points on the boundary of the tetrahedron,
conditions (\ref{lastfound}) and
(\ref{beta6}) uniquely determine
values for $\beta_5$ and $\beta_6$ from
the values of $\beta_1$, $\beta_2$, and $\beta_3$.
For points in the interior of the tetrahedron,
conditions (\ref{lastfound}) and (\ref{beta6}) give
the allowable values for $\beta_5$ and $\beta_6$
as a deformed circle, shown in Figure~\ref{b56fig}.

Finally, let us consider entanglement types for which $\beta_4 = 1/4$.
In this case, conditions (\ref{lastfound}) and (\ref{bottomsurface})
lead to requirements that can only be satisfied if
$\beta_1 = \beta_2 = \beta_3 = \beta_5 = \beta_6 = 0$.
Thus, there is just a single type of entanglement with $\beta_4 = 1/4$,
and it is the entanglement type of the GHZ state.

Table~\ref{tabtypes} classifies three-qubit entanglement types
in a topological way,
viewing $\cale_3$ as a 5-dimensional cell complex.
Each entanglement type belongs to exactly one cell
(one row of the table).
In addition to listing the ranges of the invariants for each cell
of entanglement types,
we also give the classification of \acin \etal, the SLOCC class,
and the orbit dimension for entanglement types in each cell.

\acin \etal\ classified entanglement for pure three-qubit states in
\cite{acin00}.  Their classification was based on the number of basis
states in (\ref{stdform}) required to express a standard state
with a particular entanglement type.
In Table~\ref{tabtypes}, we list the \acin type from~\cite{acin00}
associated with each cell of LU entanglement types.

The fourth column of Table~\ref{tabtypes} lists the SLOCC class
for each cell of LU entanglement types.
In the case of pure
three-qubit states, the number of SLOCC entanglement types is
finite:  there are six~\cite{dur00}.

In addition to the classification scheme of \acin and the
classification based on SLOCC, entanglement types can be classified by
the dimension of their local unitary orbits.  Each LU orbit is a
differentiable manifold with a well-defined dimension.  Carteret and
Sudbery found the orbit dimensions for three-qubit pure states in
\cite{carteret00a}.  The analysis of orbit dimensions appears to be a
promising route toward classification of $n$-qubit
states~\cite{lyonswalck1,lyonswalck2}.  In the last column of
Table~\ref{tabtypes}, we give the orbit dimension calculated for
entanglement types in each cell.

\begin{table*}
\caption{\label{tabtypes}
Three-qubit space of entanglement types viewed as a cell complex.
The superscript in the cell name gives the dimension of the cell.
The function $F$ is defined in (\ref{fdef}).}
\begin{ruledtabular}
\begin{tabular}{lllcc}
Cell & Invariant & \acin              & SLOCC              & Orbit \\
name & ranges   & Type~\cite{acin00} & Class~\cite{dur00} & Dimension \\ \hline
 & $\beta_4 = 0$ (consequently $\beta_5 = 2 \sqrt{\beta_1 \beta_2 \beta_3}$,
    $\beta_6 = 0$) &&& \\
$e^0_{A-B-C}$ & \spc $\beta_1 = \beta_2 = \beta_3 = 0$ & 1 & $A-B-C$ & 6 \\
$e^1_{A-BC}$  & \spc $0 < \beta_1 < 1/4$, $\beta_2 = \beta_3 = 0$
  & 2a & $A-BC$ & 7 \\
$e^0_{A-BC}$  & \spc $\beta_1 = 1/4$, $\beta_2 = \beta_3 = 0$
  & 2a & $A-BC$ & 5 \\
$e^1_{B-AC}$  & \spc $0 < \beta_2 < 1/4$, $\beta_1 = \beta_3 = 0$
  & 2a & $B-AC$ & 7 \\
$e^0_{B-AC}$ & \spc $\beta_2 = 1/4$, $\beta_1 = \beta_3 = 0$
  & 2a & $B-AC$ & 5 \\
$e^1_{C-AB}$ & \spc $0 < \beta_3 < 1/4$, $\beta_1 = \beta_2 = 0$
  & 2a & $C-AB$ & 7 \\
$e^0_{C-AB}$ & \spc $\beta_3 = 1/4$, $\beta_1 = \beta_2 = 0$
  & 2a & $C-AB$ & 5 \\
$e^3_W$ & \spc $0 < \beta_1 \beta_2 + \beta_1 \beta_3 + \beta_2 \beta_3
                < \sqrt{\beta_1 \beta_2 \beta_3}$
  & 4a & $W$ & 9 \\
$e^2_W$ & \spc $0 < \beta_1 \beta_2 + \beta_1 \beta_3 + \beta_2 \beta_3
                = \sqrt{\beta_1 \beta_2 \beta_3}$
  & 3a & $W$ & 8 \\
& $0 < \beta_4 < 1/4$ &&& \\
$e^1_{GHZ}$ & \spc $\beta_1 = \beta_2 = \beta_3 = 0$
 & 2b & $GHZ$ & 7 \\
$e^1_{A,GHZ}$ & \spc $\beta_1 = 1/4 - \beta_4$, $\beta_2 = \beta_3 = 0$
 & 3b & $GHZ$ & 8 \\
$e^1_{B,GHZ}$ & \spc $\beta_2 = 1/4 - \beta_4$, $\beta_1 = \beta_3 = 0$
 & 3b & $GHZ$ & 8 \\
$e^1_{C,GHZ}$ & \spc $\beta_3 = 1/4 - \beta_4$, $\beta_1 = \beta_2 = 0$
 & 3b & $GHZ$ & 8 \\
$e^2_{A,GHZ}$ & \spc $0 < \beta_1 < 1/4 - \beta_4$, $\beta_2 = \beta_3 = 0$
 & 3b & $GHZ$ & 8 \\
$e^2_{B,GHZ}$ & \spc $0 < \beta_2 < 1/4 - \beta_4$, $\beta_1 = \beta_3 = 0$
 & 3b & $GHZ$ & 8 \\
$e^2_{C,GHZ}$ & \spc $0 < \beta_3 < 1/4 - \beta_4$, $\beta_1 = \beta_2 = 0$
 & 3b & $GHZ$ & 8 \\
$e^2_{BC}$ & \spc $\beta_1 = 0$, $\beta_2 \beta_3 > 0$, $F=0$
 & 5  & $GHZ$ & 9 \\
$e^2_{AC}$ & \spc $\beta_2 = 0$, $\beta_1 \beta_3 > 0$, $F=0$
 & 4b & $GHZ$ & 9 \\
$e^2_{AB}$ & \spc $\beta_3 = 0$, $\beta_1 \beta_2 > 0$, $F=0$
 & 4b & $GHZ$ & 9 \\
$e^3_{BC}$ & \spc $\beta_1 = 0$, $\beta_2 \beta_3 > 0$, $F>0$
 & 5  & $GHZ$ & 9 \\
$e^3_{AC}$ & \spc $\beta_2 = 0$, $\beta_1 \beta_3 > 0$, $F>0$
 & 4b & $GHZ$ & 9 \\
$e^3_{AB}$ & \spc $\beta_3 = 0$, $\beta_1 \beta_2 > 0$, $F>0$
 & 4b & $GHZ$ & 9 \\
$e^3_{ABC}$ & \spc $\beta_1 \beta_2 \beta_3 > 0$, $F = 0$
 & 5 & $GHZ$ & 9 \\
$e^4$ & \spc $\beta_1 \beta_2 \beta_3 > 0$, $F > 0$,
   $\beta_5 = 2 \sqrt{\beta_1 \beta_2 \beta_3}$
 & 5 & $GHZ$ & 9 \\
$e^5$ & \spc $\beta_1 \beta_2 \beta_3 > 0$, $F > 0$,
   $\beta_5 < 2 \sqrt{\beta_1 \beta_2 \beta_3}$
 & 5 & $GHZ$ & 9 \\
& $\beta_4 = 1/4$ &&& \\
$e^0_{GHZ}$
 & \spc (consequently $\beta_1 = \beta_2 = \beta_3 = \beta_5 = \beta_6 = 0$)
 & 2b & $GHZ$ & 7 \\
\end{tabular}
\end{ruledtabular}
\end{table*}

\section{Topology and Standard States}
\label{topss}

The Schmidt decomposition theorem provides a set of standard quantum states
and a claim that every bipartite quantum state is LU equivalent to
one of the standard states.
Various generalizations of the Schmidt decomposition
to multipartite systems have been proposed
\cite{acin00,carteret00}.
Standard states, such as (\ref{stdform}), have been invaluable
in understanding three-qubit entanglement.
The two-qubit Schmidt decomposition theorem, which states that
every two-qubit state is LU-equivalent to a state of the form
$\alpha_1 \ket{00} + \alpha_2 \ket{11}$, with $\alpha_1$ and $\alpha_2$
real, $\alpha_1 \geq \alpha_2$, and $\alpha_1^2 + \alpha_2^2 = 1$,
is particularly nice in that it provides
\begin{itemize}
\item exactly one standard state for each LU entanglement type, and
\item standard states that are close together when their entanglement
types are close together.
\end{itemize}

Generalizations of the Schmidt decomposition to three qubits or more
have not been able to retain both of these properties.
This naturally leads to the question of whether a nice set of
three-qubit standard states exists, and has yet to be found,
or does not exist.

We use
topological properties of $\cale_3$ to argue that no
nice set of standard states exists for three qubits.
We can translate the two nice properties above into
mathematical requirements.
The first property, that each entanglement type have a unique standard
state, is imposed by requiring a map from the space of entanglement
types to the space of quantum states.
For three qubits,
we need a map $\sigma: \cale_3 \to \mathbb{CP}^7$.
The second property, that entanglement types that are close have
standard states that are close, is satisfied by requiring that
the map $\sigma$ be continuous.

Recall that we already have a continuous projection map
$\pi: \mathbb{CP}^7 \to \cale_3$.  The composition of these two,
$\pi \circ \sigma$, must be the identity map on $\cale_3$.
\[
\cale_3 \xrightarrow{\sigma} \mathbb{CP}^7 \xrightarrow{\pi} \cale_3
\]

Continuous maps between topological spaces induce homomorphisms
between the associated homology groups.
We will argue below that $\cale_3$ has the homotopy type of $S^5$,
in which case
the homology group $H_5(\cale_3) = H_5(S^5)$ is not trivial.
On the other hand, $H_5(\mathbb{CP}^7)$ and $H_5(S^{15})$
are trivial~\cite{hatcher}.
If there were a continuous map $\sigma: \cale_3 \to \mathbb{CP}^7$,
then in the composition of homomorphisms
\[
H_5(\cale_3) \xrightarrow{\sigma_*} H_5(\mathbb{CP}^7)
 \xrightarrow{\pi_*} H_5(\cale_3) ,
\]
$\pi_* \circ \sigma_*$ must be the identity map on $H_5(\cale_3)$.
But this is impossible, since $H_5(\mathbb{CP}^7)$ is trivial.
It follows that there is no continuous map
$\sigma: \cale_3 \to \mathbb{CP}^7$ such that 
$\pi \circ \sigma$ is the identity map on $\cale_3$.
Hence there is no set of three-qubit standard states
with the two nice properties above.

It remains to argue that $\cale_3$ has the homotopy type of $S^5$.
For a fixed value of $\beta_4$ in the range $0 < \beta_4 < 1/4$,
in Section \ref{vissec} we found a deformed tetrahedron of possible
values of $(\beta_1,\beta_2,\beta_3)$, with an additional
circle of possibilities for $(\beta_5,\beta_6)$ when
$(\beta_1,\beta_2,\beta_3)$ was in the interior of the tetrahedron.
A tetrahedron is homeomorphic to a ball
$D^3 = \{(x,y,z) \in \mathbb{R}^3 | x^2 + y^2 + z^2 \leq 1\}$.
The additional circle of possibilities in the interior of the
ball, but not on the boundary of the ball implies
that the entire
space of entanglement types with a fixed $\beta_4$ is homeomorphic
to the quotient space of $D^3 \times S^1$ in which $S^1$ collapses to
a point on the boundary of $D^3$.
The space of entanglement types with a fixed $\beta_4 \in (0,1/4)$
is therefore homeomorphic to $S^4$.
If this were the case for all values of $\beta_4$, including $0$ and
$1/4$, then the space of entanglement types would look like
$[0,1/4] \times S^4$.  But we know that when $\beta_4 = 1/4$,
the 4-sphere shrinks to a point, and when $\beta_4 = 0$,
the 4-sphere collapses to $D^3$.
We conclude that $\cale_3$ is homeomorphic to the quotient space
of $[0,1/4] \times S^4$ in which $S^4$ collapses to $D^3$ at the left
endpoint of $[0,1/4]$, and $S^4$ collapses to a point at the right
endpoint.

Now, $S^5$ is homeomorphic to the quotient space of
$[0,1/4] \times S^4$ when $S^4$ shrinks to a point at both ends of $[0,1/4]$.
The quotient map $\cale_3 \to S^5$ that collapses
$D^3$ to a point is a homotopy equivalence since
$(\cale_3,D^3)$ is a CW pair~\cite[p. 11]{hatcher}.
We conclude that $\cale_3$ and $S^5$ have the same homotopy type.

Note that $\cale_2$, by comparison,
has trivial topology
(the closed interval $[0,1]$ is contractible),
which explains why the two-qubit
Schmidt decomposition can provide a nice set of standard states.

\section{Acknowledgments}

The authors thank Research Corporation for their support of this work.
We also thank David Lyons for reading the manuscript.

\end{document}